\documentclass[twocolumn,prb,showpacs,preprintnumbers,amsmath,amssymb]{revtex4} 
\usepackage{graphicx}
\usepackage{dcolumn}
\input epsf

\begin{document}

\title{Optical conductivity of Ortho-II YBa$_2$Cu$_3$O$_{6.5}$}
\author{E. Bascones$^1$, T. M. Rice$^1$,  A.O. Shorikov$^2$,
  V.I. Anisimov$^2$}
\affiliation{$^1$Theoretische Physik, ETH-H\"onggerberg, CH-8050 Zurich (Switzerland)\\
$^2$Institute of Metal Physics, 620219 Ekaterinburg, GSP-170 (Russia)}
\date{\today}
\begin{abstract}
The Ortho-II phase of YBa$_2$Cu$_3$0$_{6.5}$ is characterized by a periodic 
alternation of  empty Cu and  filled Cu-O b-axis chains doubling 
the unit cell in the a direction. The extra oxygen in the full chains 
gives rise to an attractive potential for the holes in the planes. The planar bands split in two with a gap opening at the new 
Brillouin zone boundary $k_x=\pm \pi/2$, which we estimate from LDA 
calculations. Using a planar model which treats the d-wave superconductivity 
in a mean field approximation, we show that interband transitions produce a 
strongly anisotropic feature in the optical conductivity controlled by a 
region in $\vec{k}$-space close to $(\pi/2,\pi/2)$. The edge position of this feature 
gives information on the temperature dependence of quasiparticle spectrum
in this region. Bilayer 
splitting would show up as a double edge shape.
\end{abstract}
\pacs{74.72-h,74.25,78.30.-J,71.27+a}
\maketitle

Highly ordered Ortho-II YBa$_2$Cu$_3$O$_{6.5}$, with an average doping per planar Cu $x \sim 0.1$ and a 
T$_c \sim 60$ K, is a suitable system to 
study the properties of underdoped cuprates and the effect of the chains on
the superconducting planes in YBCO. The Ortho-II phase is characterized by a periodic 
alternation of filled and empty Cu-O b-axis chains, doubling the size 
of the unit cell in the a- direction\cite{Andersen99}. The oxygen ordering reduces the disorder and highly 
ordered samples have been prepared\cite{Liang2000}. 
Unfortunately YBCO is not suitable for the angle-resolved photoemission (ARPES) measurements\cite{Shenreview} due to poor quality 
of the cleaved surface, but it is a convenient system for
infrared experiments. 

In this letter we analyze how the oxygen ordering affects the band structure 
and optical conductivity of Ortho-II YBa$_2$Cu$_3$O$_{6.5}$. 
The presence of oxygen in the chains gives rise to an attractive potential for
the holes in the planes. Due to the alternation of empty and filled chains,
this potential is $2a$ periodic. This causes a reduction of the Brillouin zone 
in the $GX$ direction (see notation in Fig. 1) with the corresponding splitting of the 
bands. We find  a highly anisotropic optical conductivity due to interband transitions 
between the split bands, 
characterized 
by a  sharp-edge peak in the a direction, while it vanishes at 
threshold in 
the b direction. These features 
should be absent in non-Ortho II phases. The conductivity onset of absorption
is controlled by a region in $\vec{k}$-space close to $(\pi/2,\pi/2)$ and the
position of the threshold depends on the quasiparticle dispersion in this
region and on the value of the interband gap.

The normal state of underdoped cuprates has an anisotropic gap, with 
an overall d-wave dependence.
ARPES measurements suggest\cite{Norman98} that with decreasing temperature, 
the 
opening of the gap removes larger portions of the Fermi surface (FS) resulting
in Fermi arcs close to the nodal direction.  
The size of these arcs shrink to a 
point upon entering the superconducting state. 
In a bilayer material interplanar coherence appears 
as a splitting of the planar bands, and consequently of the FS branches, 
into bonding (B) and antibonding (A) states 
$\epsilon_{A,B}({\bf k})=\epsilon_{pl}({\bf k}) \pm t_\perp ({\bf k})$.
The existence of bilayer splitting (BS) in the antinodal region in the
overdoped regime is widely accepted\cite{Feng2001,Chuang2001PRL}. On the
contrary, results in the nodal direction\cite{Andersen95,Kordyuk2003condmat}, as well as
in the underdoped and optimally doped regimes\cite{Chuang2001condmat,Kordyuk2002,Borisenko2002,Borisenko2003PRL, 
Kordyuk2003condmat,Borisenko2003condmat,Kaminski03} remain controversial.
We show that the dependence of the Ortho-II interband transition threshold 
on the gap amplitude allows one  to 
extract information on these questions from infrared measurements.

We have performed LDA\cite{Kohn99} calculations using the TB-LMTO-ASA computation scheme
 \cite{Andersen84}. YBa$_2$Cu$_3$0$_{6.5}$ crystallizes in orthorhombic structure (space
grout Pmmm).  The following lattice parameters obtained at room temperature\cite{Grybos94}
 a=7.659$\AA$ b=3.872$\AA$ and c=11.725$\AA$ were used in calculations
The basis of muffin-tin orbitals contained (3s,2p,3d) orbitals for O,
(4s,4p,3d) orbitals for Cu, (5s,5p,4d,4f) orbitals for Y and (6s,6p,5d,4f)
orbitals for Ba atom. A k-point mesh of 6x4x2 was used for self consistency
calculation loop.
As suggested by recent NMR experiments\cite{Hardynmr}, the $2a$ periodicity arising from the Ortho-II ordering 
produces a charge imbalance between the Cu atoms
below a filled and an empty chain. 
Evidence of charge modulation on the order of 0.05 in planar Cu has been found
in resonant soft X-ray scattering rate\cite{Sawatzky}. We have calculated the
charge distribution in the Cu atoms using the same size for all Cu
spheres. Our estimation for the charge modulation in Cu ions (including
the occupation of all the orbitals in the basis) is on the order of 0.06 holes
in the planes.

\begin{figure}
 \begin{center}
        \leavevmode
        \epsfxsize=75mm 
        \epsfbox{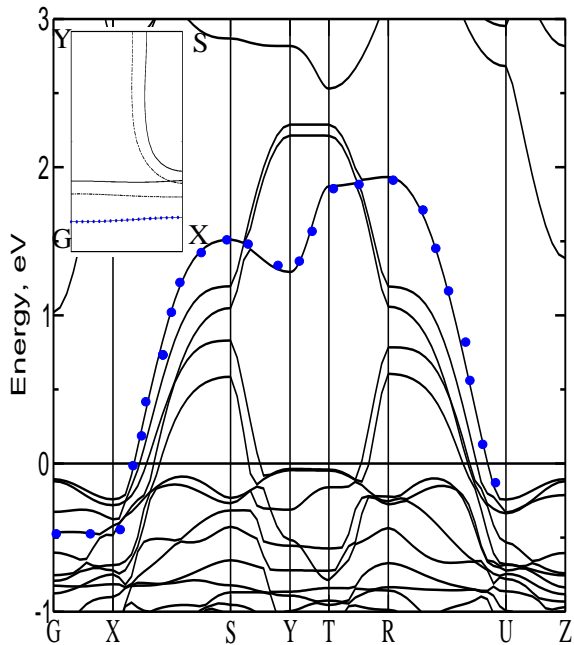}
 \end{center}
 \vspace{0mm}
 \caption{Main figure: Band structure obtained from LDA
   calculations. The band marked with dots is chain-like. $G=(0,0,0)$, 
$X=(\pi/2,0,0)$, $S=(\pi/2,\pi,0)$,
   $Y=(0,\pi,0)$, $T=(0,\pi,\pi)$, $R=(\pi/2,\pi,\pi)$, $U=(\pi,0,\pi)$, 
$Z=(0,0,\pi)$.Inset: LDA Fermi surface in the GXSY plane. The chain branch is
   enhanced with dots. Solid (dot-dashed) lines correspond to the split B (A) 
bands.}
\label{FIG. 1.}
 \end{figure} 
The calculated band structure is shown in Fig. 1. The band marked with dots is
the single chain band which crosses the Fermi level, associated with the
filled chain. Rough estimate of the volume enclosed by the Fermi surface of
the chain bands leads to $0.04 \pm 0.01$ holes per Cu in the planes. This
value is smaller than expected, what could be due to deviations of LDA
predictions close to the insulating phase.
The effect of the $2a$ periodicity is reflected 
in the band structure. 
The planar B and
A states each split in two bands.
The splitting of the bands 
shows up clearly in the XS and RU directions, at which the Brillouin zone has folded. 
The four planar bands appear in two pairs. Intra and inter-pair splitting are
due to interlayer coherence and oxygen ordering respectively.
Without the doubling of the unit cell both band pairs would be degenerate.
The interband transitions discussed below connect lower ($\alpha$)
and upper ($\beta$) band states with the same bilayer symmetry. 
LDA calculations show a $k$-dependent  $\alpha (\beta)$ splitting. On the
Fermi level, in the XS direction, it varies between 120 and 160 meV
independent of the bilayer band. 
The FS is shown as an inset. 
The chain branch is enhanced with dots. B and A branches are in solid and
dashed-dot lines respectively. 
$\beta$ bands have quasi-onedimensional
character. This quasi-onedimensionality has been used to explain the
anisotropy of the elastic scattering rate of Ortho-II $YBa_2Cu_3O_{6.5}$\cite{Harris}.  
B 
and A bands cut the XS line at $B\alpha$, $B\beta$ and $A\alpha$, $A\beta$ 
respectively. Note that BS does not vanish at the nodes in LDA.

To calculate the optical conductivity we use a planar dispersion
\begin{eqnarray}
\epsilon({\bf k})=& - & 2t(\cos k_x + \cos k_y) +  4t' \cos k_x\cos k_y\nonumber   \\  
& - &
2t''(\cos 2k_x + \cos 2k_y)-\mu  
\label{disp1}
\end{eqnarray}
$k_x,k_y \in (-\pi,\pi)$ in lattice parameter units, which we take
equal for a and b axes. 
The alternation of empty and filled chains is
introduced by an external potential
$V \sum_{i,j,P=1,2,\sigma}c^{P
  \dagger}_{2i,j,\sigma}c^P_{2i,j,\sigma}$
where $i,j$ label sites in a and b direction, $P$ the two superconducting
planes. The interlayer hopping is
$t_\perp({\bf k})$.   
In the reduced Brillouin zone $k_x \in
(-\pi/2,\pi/2)$, $k_y \in (-\pi,\pi)$ the resulting dispersion is
\begin{eqnarray}
\epsilon^{L}_{\alpha,\beta}({\bf k})=&-&2t\cos k_y - 2 t''(\cos 2k_x+ \cos 2k_y) -
\mu   \pm  t_\perp({\bf k})\nonumber \\ & \pm & \left (4 \cos^2 k_x (t - 2t'\cos k_y)^2 +
  \frac{V^2}{4} \right )^{1/2}
\end{eqnarray}
First plus and minus signs correspond to $L=A$ and $B$ and second signs to
$\alpha$ and $\beta$, respectively. Since in LDA calculations $V$ and
$t_\perp ({\bf k})$ are only slightly $\vec{k}$-dependent near the nodes, we set them constant.

To describe the d-wave superconducting state we introduce a phenomelogical
attractive potential of the form
$V_0(\cos k_x - \cos k_y)(\cos k_x'-\cos k_y ')$ which reflects the symmetry
of the underlying Cu-O plaquette. 
Thermal\cite
{Sutherland03} and microwave conductivity\cite{Hardymw} measurements 
performed in Ortho-II $YBa_2Cu_3O_{6.5}$  have 
found that its superconducting quasiparticles are well defined BCS excitations 
corresponding to an anisotropic low-energy gap with nodal lines. 
In the following, superconductivity is treated at the mean
field level in the new split bands. Only intraband and interband pair
scattering processes $(\gamma,{\bf k}, \sigma; \gamma, -{\bf k}, \sigma)
\rightarrow (\gamma ',{\bf k}',\sigma ';\gamma', -{\bf k}',-\sigma')$, with
$\gamma,\gamma '=\alpha,\beta$ are
kept. The presence of the $2a$ periodic potential breaks the rotation symmetry
of the d-wave gap and, in the reduced Brillouin zone, it takes the form:
$\Delta_{\alpha,\beta}({\bf k})=\frac{1}{2}\Delta_0(g({\bf k})\cos k_x - \cos
k_y)$ with
\begin{equation}
g({\bf k})=\frac{2 \cos k_x (t - 2 t' \cos k_y)}{\left ( 4 \cos^2 k_x (t - 2t'
    \cos k_y)^2 + \frac{V^2}{4}\right )^{1/2}}
\end{equation}
The breaking of the tetragonal symmetry shifts slightly the position of the
nodes, with a maximum shift at $k_x=0$. 
Here we take
$\Delta_0= 71$ meV for both B and A bands\cite{Feng2001,Borisenko2002,Sutherland03}.

In linear response the real part of the optical conductivity reads
\begin{widetext}
\begin{eqnarray}
\sigma^{ii}(\omega)= \frac{2 e^2}{\nu \omega} \sum_{{\bf
    k},L} M^{ii}({\bf k})\left [\frac{1}{2}\left(1 - \frac{\epsilon^L_{\alpha}({\bf
    k})\epsilon^L_{\beta}({\bf k}) + \Delta_\alpha({\bf k})\Delta_\beta({\bf
    k})}{E^L_\alpha({\bf k})E^L_\beta({\bf k})}\right)\left[1 - n_F \left (E^L_\alpha({\bf
    k})\right ) -n_F \left (E^L_\beta({\bf k})\right )\right]\delta\left(\omega - \left ( E^L_\alpha({\bf
    k}) + E^L_\beta ({\bf k})\right ) \right)\right. \nonumber \\ +  \left. \frac{1}{2}\left(1 + \frac{\epsilon^L_{\alpha}({\bf
    k})\epsilon^L_{\beta}({\bf k}) + \Delta_\alpha({\bf k})\Delta_\beta({\bf
    k})}{E^L_\alpha({\bf k})E^L_\beta({\bf
    k})}\right)\left[n_F\left(E^L_\alpha({\bf k})\right
    )-n_F\left(E^L_\beta({\bf k})\right )\right
    ]\delta \left( \omega - \left | E^L_\beta({\bf k})-E^L_\alpha({\bf k})\right | \right )
\right]   
\label{cond}
\end{eqnarray}
\end{widetext}

Here, $\nu$ and $e$ are the volume of the sample and the electronic charge,
$E^L_{\alpha,\beta}=\sqrt{\epsilon^{L 2}_{\alpha,\beta}({\bf k})+
  \Delta^2_{\alpha,\beta}({\bf k})}$ are the
quasiparticle energies in the superconducting state, $n_F$ is the Fermi
distribution function and $M^{ii}({\bf k})$ the
matrix elements in the a and b directions. 
\begin{eqnarray}
M^{aa}({\bf k})= \frac{V^2(t - 2t' \cos k_y)^2\sin ^2 k_x}{4 \cos^2 k_x(t -
  2t' \cos k_y)^2 + \frac{V^2}{4}} \\
M^{bb}{(\bf k})=\frac{4 V^2 t'^2 \sin^2 k_y \cos^2 k_x}{4 \cos^2 k_x(t -
  2t' \cos k_y)^2 + \frac{V^2}{4}}
\end{eqnarray}
The terms in parenthesis 
in (\ref{cond}) 
are the well known coherence factors.
When $\Delta_{\alpha}({\bf k})=\Delta_\beta({\bf k})=0$ the first (second)
factor vanishes if the states involved in the transition are on the same
(different) side of the FS. 
The first term in (\ref{cond}) accounts for the contribution of those
transitions which produce an excitation in both the $\alpha$ and $\beta$ bands. The second term includes
transitions which transfer a thermal excitation from one
band to another. 
\begin{figure}
 \begin{center}
        \leavevmode
        \epsfxsize=85mm 
        \epsfbox{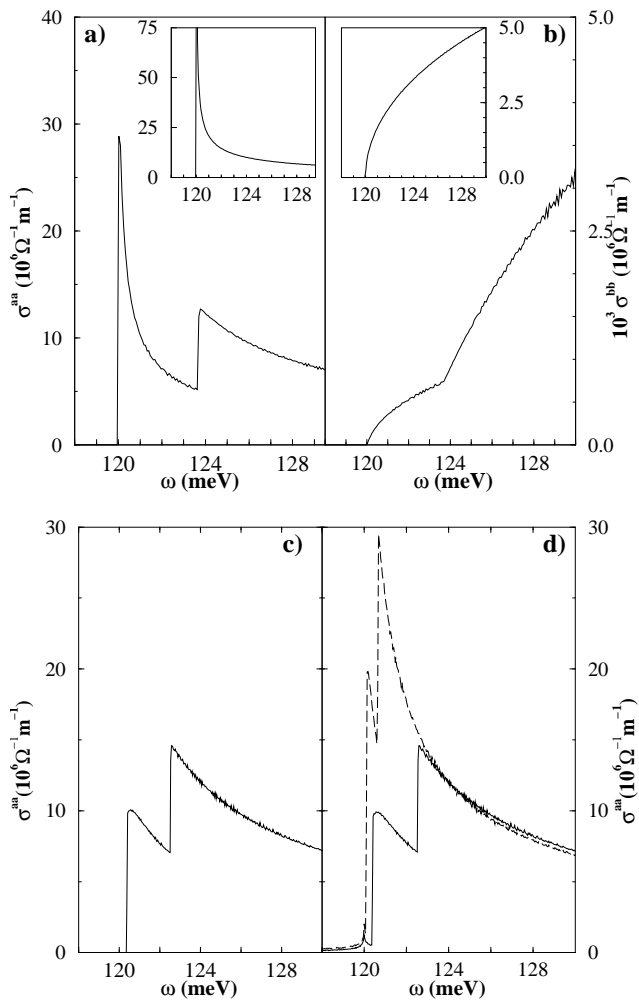}
 \end{center}
 \vspace{0mm}
\caption{Figs. a) and b) Optical conductivity in
   the a and b directions, respectively,   in the normal gapless case (insets)
   and in the superconducting state (main figures) for the 
parameters given in the text
   that 
    mimic the
 LDA calculations. Note the different scale
   in the axis of a) and b). 
The axis and units of each inset are equal to
   the ones in the corresponding figure. Figures c) and d) show the
the optical conductivity in the a direction in the superconducting state
for the renormalized dispersion: $t=180$ meV,
   $\mu=-156$ meV and $t_\perp=26$ meV, $t'/t$, $t''/t$ and
   $V$ are kept as a), see  text. The values of $\mu$ and $t_\perp$
   correspond to a doping $x_A=0.15$ and $x_B=0.05$ holes in the
   A and B bands, respectively. c) shows the zero temperature conductivity
   for  $\Delta_0=71$ meV. In d) the conductivity 
   at $60$ K corresponding to $\Delta_0=71$ meV (solid line) and
   to $\Delta_0=35.5$ meV (dashed line) is plotted.}
\label{FIG. 2.}
 \end{figure}

Upper part of Fig. 2 show the optical conductivity in the a and b directions
for parameters which mimic reasonably well the LDA bands and FS in the GXSY
plane. We take $t=360$ meV, 
$t'/t=0.3$, $t''/t=0.15$, $V=120$ meV
$\mu=-243$ meV and $t_\perp = 135$ meV. 
Fig. 2a) show the a-axis conductivity in the superconducting (main figure) and
normal, gapless, case (inset) at zero temperature. 
In the normal state the conductivity is characterized by a sharp single peak
with an onset energy  $\omega=120$ meV, while in the superconducting state it shows a double peak.
At $T=0$ there are no excitations in the system and the second
term in (\ref{cond}) vanishes. Only those transitions which cross  $E_F$ are allowed. The
minimum energy for the transition, and consequently the onset for absorption, 
is $\omega=V=120$ meV\cite{splitting} and corresponds to the states with
$k_x=\pm \pi/2$ between the two branches of the same bilayer FS, i.e. states
between $A\alpha$ and $A\beta$, and between $B\alpha$ and $B\beta$ (and the corresponding
ones in the other quadrants) in the inset of Fig. 1. Note the vicinity of
these points to the nodal direction.
Only these states contribute to the conductivity at threshold. 

A finite superconducting gap increases the energy of the
transitions. 
The minimum energy still corresponds to a state at the edge
of the zone, between the two normal state 
FS branches. 
The double peak lineshape in the superconducting state marks the
threshold energy for transitions in the B (lower onset) and A
(higher onset) bands.
The onset of interband transitions in the B-band occurs for $\vec{k}$-points
close to the nodes in the superconducting gap while in its A-band the relevant
region of $\vec{k}$-space is further away from the nodes, leading to a double
peak structure. 
The experimental observation of a double edge would be a clear evidence of the existence of BS in the
underdoped regime.

The peak lineshape is due to the $\vec{k}$ dependence of the matrix element
$M^{aa}({\bf k})$. $M^{aa}({\bf k})$ is finite and large at 
$k_x=\pm \pi/2$, and decreases away
from the zone boundary, thus the conductivity decreases with increasing
$\omega$. In the normal state, the peak in the conductivity 
is enhanced by an approximate
one-dimensional behavior of the transition energy for $\vec{k}$ between the two FS
branches close to $k_x = \pi/2$. This is also the reason of
the larger magnitude of the first peak in the conductivity in the
superconducting state. This strong enhancement would be smoothed by a $k_y$ dependent
splitting at the XS direction, as the one shown in Fig. 1, and, therefore, could be
absent experimentally.

Opposite to the behavior in the a-direction,  $M^{bb}({\bf
  k})$ vanishes at $k_x=\pm \pi/2$, resulting in a vanishing $\sigma^{bb}(\omega)$
at the threshold frequency,
as shown in Fig. 2b) for the
normal (inset) and superconducting state.  In the superconducting state the BS
  splitting would appear
in the b-axis conductivity as a double hump.
The difference in intensity in the a- and b-
directions is striking. The anisotropy in intensity and lineshape can 
be used experimentally to identify the
interband transition.

Correlation effects renormalize the band structure with
respect to the LDA predictions. In the lower part of Fig. 2 we plot the optical conductivity in
the $a$ direction, at zero (Fig. 2c) and finite (Fig. 2d) temperature, for the case in which the Fermi velocity has
been reduced to the half of its previous value. The bilayer splitting has 
been adjusted  such  that the doping in the bonding (antibonding)
band is $0.05$ holes below (above) the average doping\cite{Kordyuk2002}.
Correlations are not expected to affect the Hartree potential due to the oxygen
ordering, and we keep it unchanged.
The qualitative features in the conductivity remain. Quantitatively the peak splitting has
been slightly reduced and the bonding threshold deviates a bit from the
gapless value. 

Fig. 2d shows the optical conductivity in the a
direction $\sigma^{aa}$ at $T=60 $K.
At finite temperatures the existence of excitations reduce the contribution of
the first term in (\ref{cond}) decreasing the amplitude of the zero temperature
peak. The second term in(\ref{cond})
starts to contribute, and, in the presence of a gap,  is responsible of a small structure which appears
at $\omega\sim V$ and below. 
However, at the temperatures of interest $T\sim T_c$ these effects are very
small. On the other hand, if the superconducting gaps in the region
of $\vec{k}$-space controlling these peaks are reduced with increasing temperature,
the onsets for conduction and the peak splitting also decrease. This can be clearly seen in Fig. 2d,
where the a-axis conductivities for two different values of $\Delta_0$ are compared. The
evolution of the position of the peaks with increasing temperature gives us
information on the modification of the superconducting gap in the region of $\vec{k}$-space where
the FS cuts the $XS$ direction. 

In the pseudogap state, upon crossing $T_c$ superconducting coherence is lost,
but a gap remains in the antinodal spectrum. To describe the lose of coherence  in
(\ref{cond}) we set $\Delta_\alpha({\bf k})\Delta_\beta({\bf k})=0$ in the
coherence factors, but keep the same value (with gap) in the excitation
energies $E^L_\alpha({\bf k})$ and $E^L_\beta({\bf k})$.  This is equivalent to
the condition, in the Nambu-Gorkov formalism, that the anomalous Green functions vanish.The energy of the
interband transition does not change.
Since in the superconducting state, the values of $\Delta_\alpha({\bf k})$ and
$\Delta_\beta({\bf k})$ are small in the
area of $\vec{k}$-space which controls the conductivity close to threshold, this
modification has little effect on the
lineshape, and the discussion above remains valid even if superconducting 
coherence is lost.

In summary we predict an anisotropic feature in the infrared optical
conductivity of Ortho-II $YBa_2Cu_3O_{6.5}$, which should be absent in non
Ortho-II phases. It is a consequence of the oxygen ordering in every second chain. Possible bilayer splitting would
cause a double edge shape. The threshold energy and its evolution with
temperature give information on the quasiparticle spectrum close to the nodes
and on the effect of the chains on the planar electronic
structure. Preliminary experimental results\cite{Timusk} have not shown any
signature clearly identifiable with
the one predicted here. The estimated intensity of this feature 
is comparable with the Drude
contribution. Its appearance is based on the existence of quasiparticles. 
This picture seems to be well justified in the superconducting state
at low temperatures and low energies\cite{Sutherland03,Hardymw}.Its absence, if confirmed, could
point out to a breakdown of the quasiparticle description at higher energies.

Financial support from Swiss National Science
Foundation and NCCR MANEP and from grants RFBR-0402-16096, RFBR-03-02-39024
and URO-SO-22 is gratefully acknowledged.


\begin{thebibliography}{99}
\bibitem{splitting}{The conductivity includes the contribution of transitions in both the bonding
and antibonding states, which, as suggested by the LDA calculations have the
same onset.
If the
periodic potential seen by the bonding and antibonding bands were different, each band would
have a different optical threshold and the conductivity would show more structure, with two peaks
and a double hump in the a and b directions, respectively.}
\end{thebibliography}
\end{document}